\documentclass[amsmath,amssymb,aps,prb,reprint,superscriptaddress]{revtex4-1}
\usepackage{graphicx}% Include figure files
\usepackage{xcolor}
\usepackage[colorlinks=true,urlcolor=blue,citecolor=blue,linkcolor=blue]{hyperref}
\usepackage{physics}
\usepackage{siunitx}
\usepackage{xspace}
\usepackage{bm}
\usepackage{comment}

\newcommand{\bk}{\textbf{k}}
\newcommand{\bq}{\textbf{q}}
\newcommand{\bp}{\textbf{p}}

\newcommand{\ba}{\hat{\gamma}_\bk}
\newcommand{\bc}{\hat{\gamma}^\dag_\bk}
\newcommand{\bam}{\hat{\gamma}_{-\bk}}
\newcommand{\bcm}{\hat{\gamma}^\dag_{-\bk}}

\usepackage[normalem]{ulem}

\definecolor{JK-color}{named}{green}
\definecolor{AJ-color}{named}{blue}
\definecolor{ACG-color}{rgb}{0.97,0.57,0.11}
\definecolor{GB-color}{RGB}{128,0,128}

\definecolor{JK-color2}{named}{green}
\definecolor{AJ-color2}{named}{red}
\definecolor{ACG-color2}{rgb}{0.87,0.47,0.01}
\definecolor{GB-color2}{RGB}{128,0,128}

\begin{document}

%\preprint{}

\title{Light-induced topological superconductivity in transition metal dichalcogenide monolayers}


%\thanks{}

\newcommand{\affiliationAarhus}{Center for Complex Quantum Systems, Department of Physics and Astronomy, Aarhus University, Ny Munkegade, DK-8000 Aarhus C, Denmark}
\newcommand{\affiliationAalto}{Department of Applied Physics, Aalto University, P.O.Box 15100, 00076 Aalto, Finland}
\newcommand{\affiliationCambridge}{Departamento de F\'isica Qu\'imica, Instituto de F\'isica, Universidad Nacional Aut\'onoma de M\'exico, Apartado Postal 20-364, Ciudad de M\'exico C.P. 01000, Mexico}
\newcommand{\affiliationChina}{Shenzhen Institute for Quantum Science and Engineering and Department of Physics, Southern University of Science and Technology, Shenzhen 518055, China}

\author{Aleksi Julku}
\affiliation{\affiliationAarhus}
\author{Jami J. Kinnunen}
\affiliation{\affiliationAalto}
\author{Arturo Camacho-Guardian}
\affiliation{\affiliationCambridge}
\author{Georg M. Bruun}
\affiliation{\affiliationAarhus}
\affiliation{\affiliationChina}


\date{\today}

\begin{abstract}
Monolayer transition metal dichalcogenides (TMDs) host deeply bound  excitons interacting with itinerant electrons, and as such they  represent an exciting new quantum many-body Bose-Fermi mixture. 
Here, we demonstrate that electrons interacting with a Bose-Einstein condensate (BEC) of exciton-polaritons can  realise a   two-dimensional
 topological $p_x+ip_y$ superconductor. Using strong coupling Eliashberg theory, we show that this is caused by 
 an attractive interaction mediated by the BEC, which  overcompensates the repulsive Coulomb 
 interaction between the electrons. The hybrid light-matter nature of the BEC is crucial for achieving this, since it can be used to reduce retardation 
 effects and increase the mediated interaction in regimes important for pairing. We finally show how the great flexibility of TMDs allows one to tune the critical temperature of the topological superconducting phase to be within experimental reach.

\end{abstract}

\pacs{}
%\keywords{}

\maketitle

%\hat{x}

%%%%%%%%%%%%%%%%%%%%%%%%% 

%\paragraph{Introduction.--}.

\section{Introduction}
Atomically thin  transition metal dichalcogenides (TMDs) are an  exciting new class of truly two-dimensional (2D) semiconductors with  strong spin-orbit coupling and spin-valley locking, which provides a  rich setting for exploring new quantum states and optoelectronic applications~\cite{Wang2018,Mueller2018,Schneider2018}. Monolayer TMDs exhibit a direct band gap and optical valley selection rules~\cite{Xiao2012,Cao2012,Zeng2012,Mak2012b,Schaibley2016}, and owing to the reduced Coulomb screening they  support tightly bound excitons. Furthermore, TMDs  in optical microcavities host exciton-polaritons~\cite{Dufferwiel2015,Schneider2018,Kavokin2017} that allows to combine  the non-linear physics of the matter part with the coherence of photons~\cite{Dufferwiel2015,Schneider2018,Sidler2016,Tan2020,Emmanuele2020,Anton-Solanas2021,Bastarrachea-Magnani2019,Bastarrachea-Magnani2020,Julku2021}.

Excitons mixed with electrons in TMDs  form a new and interesting  Bose-Fermi mixtures in a solid-state setting, which compliments such mixtures realised in quantum degenerate atomic gases~\cite{Park2012,Heo2012,FerrierBarbut2014,Vaidya2015,DeSalvo2017,Lous2018,Schaefer2018}. Bose-Fermi mixtures play a key role in a diverse range of condensed matter phenomena including liquid helium, superconductivity mediated by phonons or magnons, as well as polaron physics. Experiments have so far focused on the regime of small exciton concentration where their interaction with  electrons leads to the formation of Fermi polarons~\cite{Sidler2016,Tan2020,Emmanuele2020,Efimkin2017,Shahnazaryan2020,Bastarrachea-Magnani2020}. Increasing the exciton concentration beyond the polaron regime has been predicted to give rise to a range of  intriguing effects such as trion liquids~\cite{Milczewski2022}, density ordered, and superconducting phases~\cite{Shelykh2010b,Matuszewski2012,Kovalev2011,Villegas2019,Boev2019,Laussy2010,Cotlet2016}.

%%%%%%%%%%%%%%%%%%%%%%%%%%%%%%%%%%%%%%%%%%%%%%%%%%%%%%%%%%%%%%%%%%%%%%%%%%%
\begin{figure}
  \centering
    \includegraphics[width=1.0\columnwidth]{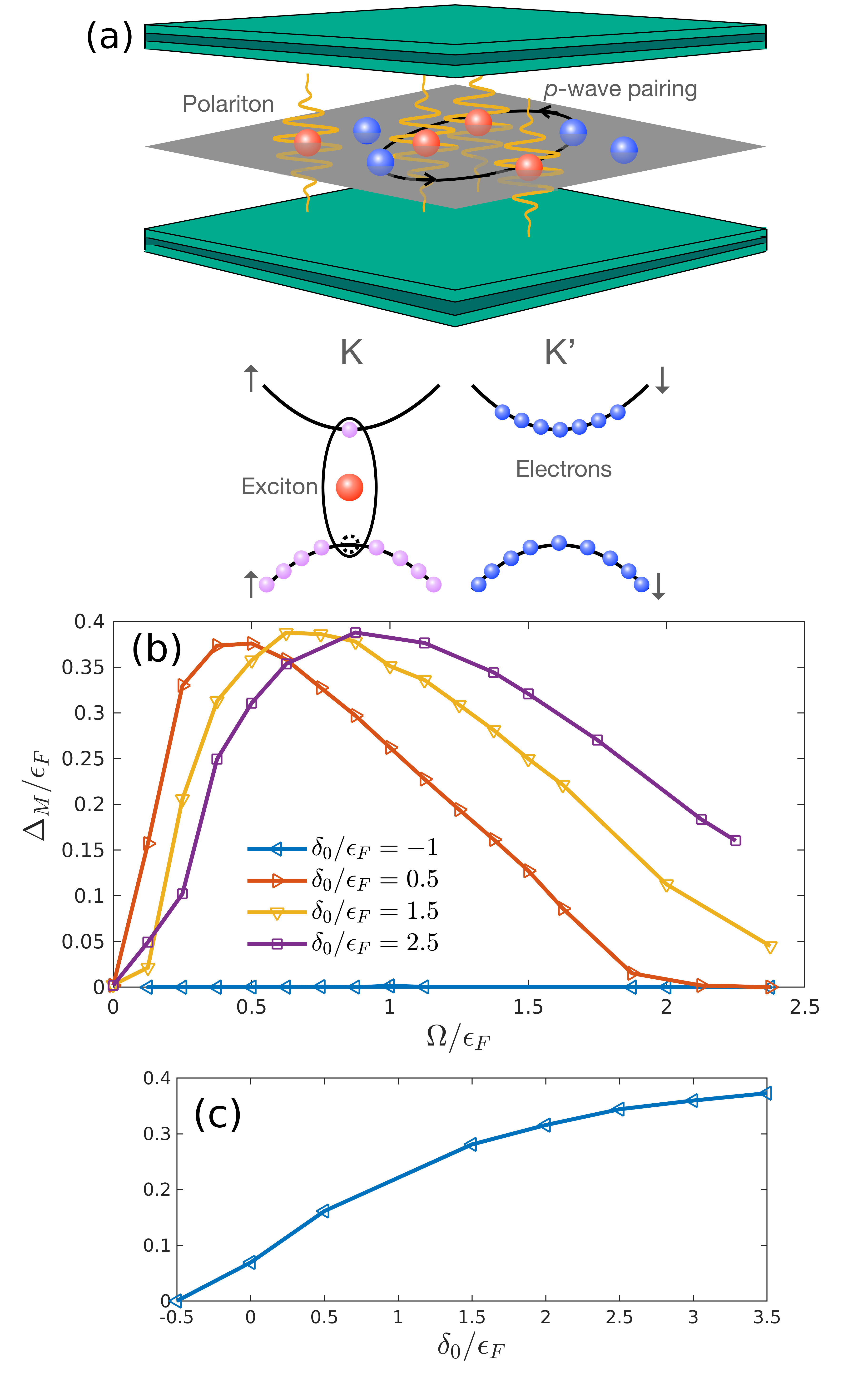}
    \caption{(a) Upper panel: Spin-polarised electrons (blue spheres) in a monolayer TMD 
    interact with polaritons formed by excitons (red spheres) hybridized with microcavity photons (yellow waves). Polaritons mediate an attractive interaction between the electrons leading to topological $p$-wave pairing (black circle). Lower panel: The excitons are created by particle-hole excitations of spin polarised electrons in the $K$-valley of the TMD  band structure, and the electrons live in  the $K'$-valley. (b) The superconducting gap $\Delta_M$ for  different values of $\delta_0$ as a function of $\Omega$. (c) $\Delta_M$  as a function of $\delta_0$ for  $\Omega/\epsilon_F = 1.375$.}
   \label{Fig:1}
\end{figure}
%%%%%%%%%%%%%%%%%%%%%%%%%%%%%%%%%%%%%%%%%%%%%%%%%%%%%%%%%%%%%%%%%%%%%%%%%%%

The role of the spin-valley physics  of TMDs have however not 
been fully explored in this context. Here, we show that the flexibility  
of the spin-valley degrees of freedom provides a promising 
platform for realising a topological superconductor. This is formed by  
spin polarized electrons electrons residing in one valley, which interact attractively 
via an induced interaction mediated by exciton-polaritons in another valley. By tuning the 
many free parameters of our setup, we show that 
the critical temperature can be optimised to be within experimental reach. 
The quest for topological superconductivity remains highly controversial in spite of 
years of intense effort~\cite{Zhang2018,*Zhang:2021ut,Vaitiekenas2020,*Sills2021,LinHe2017,*Thorp2021,Chronister2021}, and finally achieving this with our setup would  be a major breakthrough.

%\paragraph{System.--} 
\section{System}
We consider a TMD monolayer  in a microcavity where excitons are created in the $K$-valley by optical pumping. As the Coulomb screening is greatly reduced in 2D, the binding energy~\cite{Chernikov2014,Wang2018} of excitons  is so large that they can  for the present purpose be treated as  
point-like bosons~\cite{Wang2018,Efimkin2020}. The excitons   interact with $K'$-valley conduction band electrons, which are spin-polarized due to intrinsic spin-orbit coupling, 
see Fig.~\ref{Fig:1}(a). The Hamiltonian of the system is 
\begin{gather}
\hat H = \sum_{\bk} \begin{bmatrix} \hat{x}_{\bk}^\dag  \hat{c}_{\bk}^\dag \end{bmatrix}
\begin{bmatrix}
\epsilon_{\bk}^{x}  & \Omega \\ \Omega & \epsilon_{\bk}^{c} 
\end{bmatrix}
\begin{bmatrix}
\hat{x}_{\bk} \\ \hat{c}_{\bk}
\end{bmatrix} \nonumber 
 +\sum_\bk \epsilon_{\bk}^{e}  \hat{e}^\dag_\bk \hat{e}_\bk   \nonumber \\
 +\frac12\sum_{\bk,\bk',\bq} [g_{ee}(\bq) \hat{e}^\dag_{\bk'-\bq} \hat{e}^\dag_{\bk+\bq} \hat{e}_\bk \hat{e}_{\bk'} +g_{xx}(\bq) \hat{x}^\dag_{\bk'-\bq} \hat{x}^\dag_{\bk+\bq} \hat{x}_\bk \hat{x}_{\bk'}] \nonumber \\
 + \sum_{\bk,\bk',\bq} g_{ex}(\bq) \hat{x}^\dag_{\bk'-\bq} \hat{e}^\dag_{\bk+\bq} \hat{e}_\bk \hat{x}_{\bk'},
 \label{tot_ham}
\end{gather}
where $\hat{x}_\bk$, $\hat{e}_\bk$ and $\hat{c}_\bk$ annihilate a $K$-valley exciton, a $K'$-valley electron, and a cavity photon of momentum $\bk=(k_x,k_y)$ respectively. 
We assume that the electrons occupy states near the conduction band minimum so that their single particle 
dispersion $\epsilon_\bk^{e} = \bk^2/2m_e$ is quadratic, and the same argument applies for the excitons giving $\epsilon_\bk^{x} = \bk^2/2m_x$,
with $m_e$ and $m_x$ being the electron and exciton effective masses. 
The cavity gives rise to a dispersion $\epsilon_\bk^{c} = \bk^2/2m_c + \delta_0$ for the photons, where $\delta_0$ is the cavity detuning and $m_c$ the photon mass. 
%The non-interacting  energies are  $\epsilon_\bk^{x} = \bk^2/2m_x$, $\epsilon_\bk^{e} = \bk^2/2m_e$, and 
%$\epsilon_\bk^{c} = \bk^2/2m_c + \delta_0$, where $\delta_0$ is the cavity detuning , and cavity photons, . 
 We use experimentally realistic values of $m_x = 2m_e$, $m_c = 10^{-5}m_e$ and $m_e = 0.5m_0$ with $m_0$ being the bare electron mass~\cite{Kormnyos2015}. 
 Furthermore,  $\Omega$ is the exciton-photon Rabi coupling strength and 
    the second line of Eq.~\eqref{tot_ham} gives the electron-electron  and exciton-exciton interaction,  whereas the third line  is the electron-exciton interaction. 
  We use units where the system volume,  Boltzmann's, and Planck's constant  are all unity. 

  Due to the small spatial size of the excitons, the exciton-electron interaction is short range and we therefore assume a momentum independent interaction 
  $g_{ex}(\bq) = g_{ex} = 1.5$ $\mu$eV$\mu$m$^2$. This is justified by the fact that we throughout the manuscript take a small electron  density  
$n_e = k_F^2/4\pi=10^{15}$ m$^{-2}$ and $n_0/n_e= 10$, if not otherwise mentioned, with $k_F$ being the Fermi momentum and $n_0$ being the exciton density. The chosen electron density corresponds to a Fermi energy  $\epsilon_F=k_F^2/2m_e \lesssim 1$ meV. Since the binding energy of excitons is $\sim 100$ meV in TMDs~\cite{Wang2018},  the  electrons in valley $K'$ will not significantly  change the properties of the excitons in valley $K$. It has indeed been shown that for large exciton binding energies, they can to a very good approximation be treated as point bosons with a momentum independent (short range) exciton-electron interaction~\cite{Efimkin2021}, precisely as we do in the present paper.
Likewise, we take a momentum independent exciton-exciton interaction.  Recent experimental estimates range from $g_{xx}\sim 0.05\mu$eV$\mu$m$^2$~\cite{Barachati2018,Tan2020}  to $g_{xx}\sim 3.0\mu$eV$\mu$m$^2$~\cite{Emmanuele2020}, 
 and we use $g_{xx} = 0.05\mu$eV$\mu$m$^2$ in the following.  
The Coulomb interaction between the electrons is $g_{ee}(\bq) = e^2/2\epsilon q$ 
 where $\epsilon$ is the permittivity of the system and $e$ is the electron charge. We use $\epsilon = 4.5 \epsilon_0$  with $\epsilon_0$ being the dielectric constant,
  which is close to the permittivity of boron nitride, a regularly used encapsulating material.

%\paragraph{Mediated interaction.--} 
\section{Mediated interaction}
The system is in the regime of strong light-matter coupling leading to the formation of exciton-polaritons with energies 
$\epsilon^{LP/UP}_\bk = ( \epsilon_{\bk}^{c} + \epsilon_{\bk}^{x} \pm \sqrt{\delta_{\bk}^2 + 4\Omega^2})/2$, 
where $\delta_\bk = \epsilon^c_\bk -  \epsilon^{x}_\bk$~\cite{Hopfield1958}.  
Recent experiments~\cite{Anton-Solanas2021} have revealed signatures of Bose-Einstein condensation of exciton-polaritons in TMDs, and we 
 therefore consider the case where a BEC of density $n_0$ is formed in  the $\bk=0$ state of the lower polariton branch with energy $\epsilon^{LP}_{\bk=0}$. 
 Due to its large compressibility, the BEC can mediate a strong and attractive induced interaction whose dominant 
contribution is the  exchange of  sound modes in the BEC. 
%\GB{as shown in  the diagram  in Fig.~X}. 
Using Bogoliubov theory for the BEC, this yields~\cite{Camacho2021,Wu2016}
\begin{align}
\label{ind_int}
V_\text{ind}(\bk,i\omega_n) = -\frac{2n_0 (g_{ex} \mathcal{C}_0 \mathcal{C}_\bk )^2 \tilde{\epsilon}^{LP}_{\bk}}{\omega_n^2 + E^2_\bk},     
\end{align}
for the induced interaction between electrons. Here $\omega_n$ is a bosonic Matsubara frequency, $\tilde{\epsilon}^{LP}_\bk = \epsilon_\bk^{LP} - \epsilon_0^{LP}$,
 and $E_\bk = \sqrt{ \tilde{\epsilon}^{LP}_{\bk}(\tilde{\epsilon}^{LP}_{\bk} + 2g_{xx} \mathcal{C}^2_0 \mathcal{C}^2_\bk n_0) }$ is the Bogoliubov excitation energy.
  The Hopfield coefficients $\mathcal{C}^2_\bk = 1/2 + \delta_\bk/2\sqrt{\delta_\bk^2 + 4\Omega^2}$ 
  in Eq.~\eqref{ind_int} appear because it is only the excitonic component of the  polaritons that interacts with the electrons. In deriving Eq.~\eqref{ind_int}, we have assumed that the BEC density $n_0$ is much larger than that of the electrons  $n_e$, i.e. $n_0 \gg n_e$, so the BEC is largely unaffected by the  electrons. In particular, one can expect there are no roton instabilities in contrast to previous works focusing on larger electron densities~\cite{Cotlet2016,Matuszewski2012,Strashko2020,Cotlet2020}, see Appendix~\ref{roton_inst}.

%\paragraph{Light-induced pairing.--}
%\paragraph{Eliashberg theory.--}
\section{Eliashberg theory}
Equation \eqref{ind_int} shows that the induced interaction is attractive, and we will now explore whether it  can overcompensate 
the repulsive Coulomb interaction between the electrons and lead to superconductivity. 
The total electron-electron interaction is   
\begin{align}
V_\text{tot}(\bk,i\omega_n) = V_\text{ind}(\bk,i\omega_n) + \frac{ e^2}{2\epsilon q [1 + g_{ee}(\bk)\chi_0(\bk)]},   
\end{align}
where the second term is the  screened Coulomb interaction
in the random phase approximation, with 
 $\chi_0(\bk)$ being the static polarizability of the 2DEG~\cite{mahan:book}.

To explore polariton-mediated superconductivity in a reliable way, we use strong coupling Eliashberg theory by defining the matrix Green's function 
$G_{ij}(\bk,\tau)$~\cite{mahan:book,Kinnunen2018,Wu2016}
\begin{equation}
\label{greenf}
G(\bk,\tau) =- \begin{bmatrix}
   \langle T_\tau  \hat{e}_{\bk}(\tau)  \hat{e}_{\bk}^\dag(0) \rangle & \langle T_\tau  \hat{e}_{\bk}(\tau)  \hat{e}_{-\bk}(0) \rangle  \\   \langle T_\tau  \hat{e}_{-\bk}^\dag(\tau)  \hat{e}_{\bk}^\dag(0) \rangle &  \langle T_\tau  \hat{e}_{-\bk}^\dag(\tau)  \hat{e}_{-\bk}(0) \rangle
  \end{bmatrix} 
  %\nonumber \\
 % &\equiv \begin{bmatrix}
  % G_{11}(\bk,\tau) & G_{12}(\bk,\tau) \\ 
  % G_{21}(\bk,\tau) & G_{22}(\bk,\tau) \\ 
  %\end{bmatrix}.
\end{equation}
where $\tau$ is imaginary time and $T_\tau$ is the time-ordering operator. Neglecting vertex corrections~\cite{mahan:book}, 
the normal and anomalous self-energies 
are~\cite{mahan:book} 
\begin{align}
G_{11}(p) &= \frac{ip_n + \xi_\textbf{p} + \Sigma_{11}(-p)}{\det[G^{-1}(p)]}  \nonumber \\
G_{12}(p) &= \frac{\Sigma_{12}(p)}{\det[G^{-1}(p)]} \nonumber\\
 \Sigma_{ij}(p) &= -T\sum_{p'}V_\text{tot}(p-p')G_{ij}(p'),  \label{el2}%\label{el1}
 %\nonumber\\
%\Sigma_{11}(p) &= -T\sum_{p'}V_\text{tot}(p-p')G_{11}(p') \label{el2}
\end{align}
%\begin{align}
%&G_{11}(p) = \frac{ip_n + \epsilon_\textbf{p} + \Sigma_{11}(-p)}{\det[G^{-1}(p)]} \nonumber  \\
%&G_{12}(p) = \frac{\Delta(p)}{\det[G^{-1}(p)]}  \\
%&\Delta(p) = -\frac{1}{\beta A}\sum_{p'}g_{ee,tot}(p-p')G_{12}(p') \label{el1} \\
%&\Sigma_{11}(p) = -\frac{1}{\beta A}\sum_{p'}g_{ee,tot}(p-p')G_{11}(p') \label{el2}
%\end{align}
where  $p \equiv (\textbf{p},ip_n)$, $p_n$ is a fermionic Matsubara frequency, 
$T$ is the temperature, and $\xi_\textbf{p}=\epsilon_\mathbf{p}-\mu_e$ with  $\mu_e$ being  the chemical potential for the electrons.  Moreover, $G_{22}(p) = - G_{11}(-p)$ and $G_{21}(p) = G_{12}^*(p)$. The symbol $\sum_{p'}$ in Eq.~\eqref{el2} means sum over the Matsubara frequencies and integration over the momenta. We provide a more detailed discussion of these equations  in App.~\ref{AppB}.

We solve the  Eliashberg equations \eqref{el2} self-consistently  keeping the electron density $n_e = T\sum_p G_{11}(p) e^{i p_n 0^+}$ fixed by adjusting   $\mu_e$. 
A solution with a non-zero value of the gap  $\Delta(p) \equiv \Sigma_{12}(p)$ 
corresponds to the system being in a superconducting phase. 
Since we consider spin-polarized electrons, the superconducting gap $\Delta( p)$ must be anti-symmetric in momentum space. Among the different choices, the $p_x+ip_y$ symmetry is expected to have the lowest energy as it has no nodes.  When solving Eq.~\eqref{el2} numerically, we  take $\Delta(p) = \Delta_1(|\mathbf{p}|,ip_n) \exp(i\phi_\mathbf{p})$, where $\phi_\mathbf{p}$ is the polar angle of the momentum $\mathbf{p}$.
Thus, our setup naturally realises  a topological superconductor with Majorana modes at its edges.

Our approach based on solving the  Eliashberg equations should be compared to recent studies of polariton-mediated superconductivity, where the interaction is averaged over the Fermi surface to obtain a BCS-like equation~\cite{Laussy2010}, or an analytical approximation based on neglecting the momentum  dependence of the self-energies is used~\cite{Cotlet2016}. In contrast to these works, here we take into account the full frequency and momentum dependence of both the superconducting gap $\Delta( p)$ and the diagonal self-energy $\Sigma_{11}(p)$ self-consistently. As shown in Ref.~\onlinecite{Kinnunen2018}, simplified BCS-like theories might dramatically overestimate the critical temperature as compared to the full Eliashberg theory employed here. In particular, it is certainly not  guaranteed that our analysis yields a non-zero superconducting order parameter since  retardation effects in general suppress the induced interaction so that the Coulomb repulsion between the electrons may dominate.

%%%%%%%%%%%%%%%%%%%%%%%%%%%%%%%%%%%%%%%%%%%%%%%%%%%%%%%%%%%%%%%%%%%%%%%%%%%
\begin{figure}
  \centering
    \includegraphics[width=1.0\columnwidth]{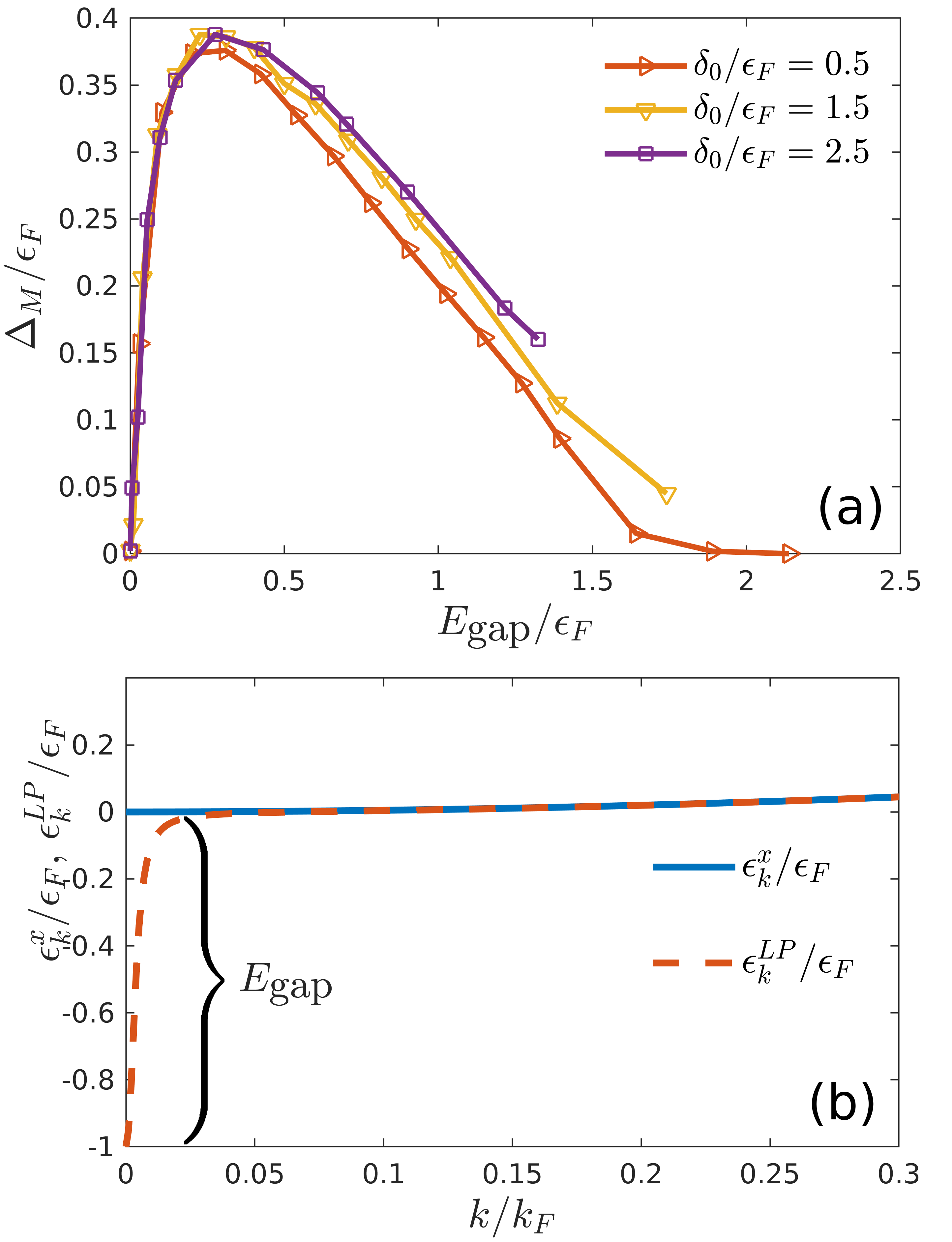}
    \caption{(a) $\Delta_M$ for different $\delta_0$ as a function of $E_{\textrm{gap}}$. (b) Exciton dispersion $\epsilon_\bk^x$ (solid blue) and a typical  exciton-polariton dispersion $\epsilon_\bk^{LP}$ (red dashed). 
    The energy difference $\epsilon_0^x - \epsilon_0^{LP}$ is denoted as $E_{\textrm{gap}}$. }
   \label{Fig:R1}
\end{figure}
%%%%%%%%%%%%%%%%%%%%%%%%%%%%%%%%%%%%%%%%%%%%%%%%%%%%%%%%%%%%%%%%%%%%%%%%%%%

%\paragraph{Results.--} 
\section{Numerical Results}
We now present results based on solving Eqs.~\eqref{el2} numerically.

\subsection{Pairing close to zero temperature}
 Let us first focus on the effects of the  cavity light as parametrised through the Rabi coupling $\Omega$ and  detuning $\delta_0$. 
In Fig.~\ref{Fig:1}(b), the maximum value of the superconducting gap at the Fermi surface,  
$\Delta_M \equiv \max_\omega \Delta(k_F,\omega)$, 
is plotted as a function of $\Omega$ for  different values of the detuning $\delta_0$ and a very low temperature $T=\epsilon_F/50$.
 We first note that for the negative detuning $\delta_0/\epsilon_F = -1$, $\Delta_M$ vanishes and there is no superconductivity.
Physically, this is a consequence of the small exciton component of the polariton condensate for $\delta_0<0$ as quantified through  the Hopfield coefficient ${\mathcal C}_0$. Since the mediated interaction is second  order in the electron-exciton interaction, this gives rise to the ${\mathcal C}_0^2$ factor in Eq.~\eqref{ind_int}, which 
 suppresses $V_{\text{ind}}$ for negative values of $\delta_0$. Figure \ref{Fig:1}(c) illustrates this further by plotting $\Delta_M$ as a function of $\delta_0$ for $\Omega/\epsilon_F = 1.375$. This shows that the gap is strongly suppressed for  $\delta_0<0$.

 Importantly, we see from Fig.~\ref{Fig:1}(b) that when the light coupling is turned on, the system becomes superconducting. Indeed, 
 while $\Delta_M \simeq 0.002 \epsilon_F$ for $\Omega=0$, which is so small that the corresponding critical temperature is well below the experimentally accessible regime, $\Delta_M$ initially increases as a function of $\Omega$,  
reaching a maximum value at some $\delta_0$-dependent value $\Omega_c >0$, after which it decreases.
To  explore why the coupling to light significantly enhances the pairing instability, we define the energy difference $E_{\text{gap}} \equiv \epsilon^{x}_0 - \epsilon^{LP}_0 = -\delta_0/2 + \sqrt{\delta_0^2/4 + \Omega^2}$ between the lowest polariton and the bare exciton at zero momentum and plot in Fig.~\ref{Fig:R1}(a) $\Delta_M$ as a function of $E_{\text{gap}}$  by varying $\Omega$ for different fixed values of $\delta_0 > 0$. This shows that different  $\delta_0$ and $\Omega$ that give the same $E_{\textrm{gap}}$ also  give approximately the same $\Delta_M$. 
Hence, the coupling to light mainly affects  the pairing  through $E_{\textrm{gap}}$ when $\delta_0>0$. To understand this, we plot in Fig.~\ref{Fig:R1}(b) the single particle dispersion $\epsilon_{\mathbf k}^\text{LP}$ 
for $\delta_0/\epsilon_F = 1/2$ and $\Omega/\epsilon_F= \sqrt{3/2}$. Due to the small photon mass $m_c$, the polariton dispersion 
is very steep around the origin and  the photons decouple for  momenta
$k\ll k_F$ %$k\gtrsim \sqrt{m_c\text{max}(\delta,\Omega)}$ 
%orders of magnitude smaller than $k_F$, 
such that the dispersion of the lower polariton approaches that  of the bare exciton with the corresponding Hopfield coefficient close to 
unity. Hence, the main effect of the light coupling  is indeed to introduce the energy gap $E_{\text{gap}}$
between the ground state polaritons forming the BEC and its excitations mediating the attractive interaction between the electrons.

\begin{figure}
  \centering
    \includegraphics[width=1.0\columnwidth]{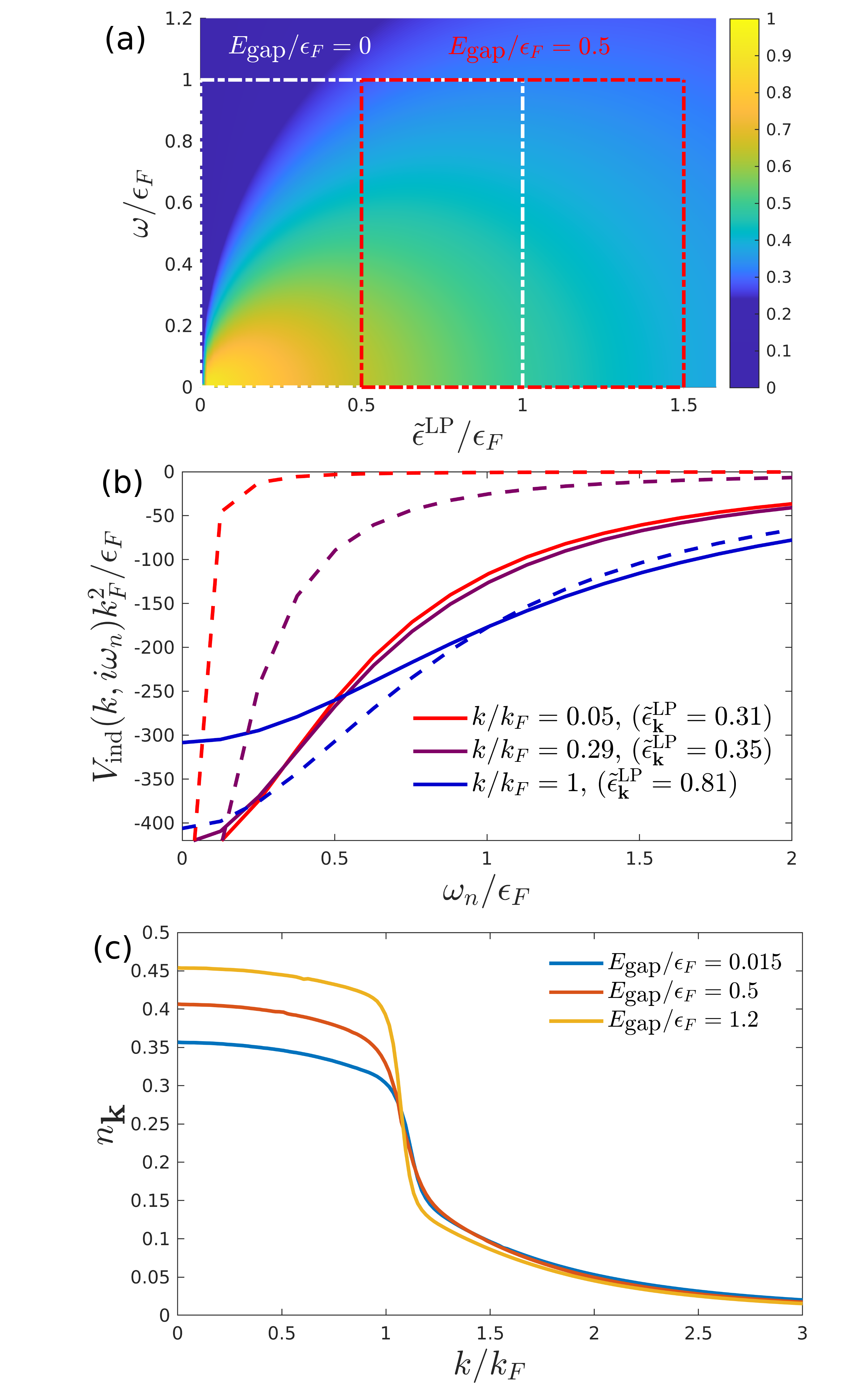}
    \caption{ (a) The induced interaction $V_{\text{ind}}$ as a function of $\tilde{\epsilon}^{\text{LP}}$ and $\omega$ for $\mathcal{C}_0\mathcal{C}_\bk \approx 1$. White and red rectangles represent the relevant regime for pairing for $E_{\text{gap}}/\epsilon_F =0$ and $E_{\text{gap}}/\epsilon_F =0.5$, respectively. (b) 
    $V_{\text{ind}}$ as a function of $\omega_n$ for different momenta in the case of excitons (dashed) and exciton-polaritons (solid) with $\delta_0/\epsilon_F = \Omega/\epsilon_F = 0.5$ ($E_{\textrm{gap}}/\epsilon_F = 0.31$). (c) Momentum distribution $n_\bk$ for  $\delta_0/\epsilon_F = 1$ and $\Omega/\epsilon_F = 0.125$ ($E_{\text{gap}}/\epsilon_F = 0.015$, blue), $\Omega/\epsilon_F =0.875$ ($E_{\text{gap}}/\epsilon_F = 0.5$, red), and $\Omega/\epsilon_F =1.625$ ($E_{\text{gap}}/\epsilon_F = 1.2$, yellow).} %(c) The effectice range $k_F/k_r$ of the pairing interaction as a function of  $E_{\textrm{gap}}$.}
   \label{Fig:R2}
\end{figure}

To further explore the role of $E_\text{gap}$, it is instructive for a moment
to approximate the polariton dispersion as $\epsilon^{LP}_\bk \approx \epsilon^{x}_\bk - E_\text{gap} \delta_{\bk,0}$ with $\delta_{\bk,0}$ being the Kronecker delta. We have checked numerically that this gives essentially the same results as using the full exciton-polariton spectrum. Figure \ref{Fig:R2}(a) plots the induced interaction $V_\text{ind}$, given in Eq.~\eqref{ind_int}, as a function of $\tilde \epsilon^\text{LP}$ and $\omega_n$ by taking $\mathcal{C}_\bk \mathcal{C}_0 \approx 1$ which is a reasonable approximation for $\delta_0>0$. The relevant momenta for superconductivity are $k\lesssim \mathcal{O}(k_F)$ corresponding to polariton energies $E_\text{gap}\lesssim \tilde \epsilon^\text{LP}_\bk\lesssim E_\text{gap}+\epsilon_F$, and  the important frequencies are likewise $|\omega|\lesssim\epsilon_F$. The corresponding region is indicated as rectangles  in Fig.~\ref{Fig:R2}(a) for $E_\text{gap}/\epsilon_F = 0$ and $E_\text{gap}/\epsilon_F = 0.5$. 
We see that the magnitude  of $V_\text{ind}$ decreases for large $\tilde \epsilon^\text{LP}_\bk$, which is easily understood by the fact that $V_\text{ind}\propto1/\tilde \epsilon^\text{LP}_\bk$. 
This  explains the suppression of superconductivity for  large $E_\text{gap}$ shown in Fig.~\ref{Fig:R1}(a). The  initial increase in the pairing with $E_\text{gap}$   is, on the other hand, caused by two effects. 

First, it follows from Eq.~\eqref{ind_int} that $V_{\text{ind}}$ becomes less dependent on $\omega$ with increasing $E_\text{gap}$. Namely, the full width at half maximum of $V_{\text{ind}}$ in the frequency space at finite  $\bk$ is given by the Bogoliubov energy, i.e. $E_\bk = \sqrt{ (\epsilon^x_\bk + E_{\textrm{gap}})^2 + 2n_0 g_{xx} (\epsilon^x_\bk + E_{\textrm{gap}}) }$ for $\mathcal{C}_\bk \mathcal{C}_0 \approx 1$. Polaritons with a non-zero $E_{\textrm{gap}}$ thus yield a broader interaction in frequency space 
suppressing retardation, which in turn  enhances Cooper pairing~\cite{mahan:book,Kinnunen2018}. 
Second, $V_\text{ind}$ increases for energies $\tilde\epsilon^\text{LP}_\bk\lesssim \omega$. Figure \ref{Fig:R2}(a) shows that these two effects make the induced interaction less dependent on frequency as well as larger in the region relevant for pairing when $E_\text{gap}\sim \epsilon_F/2$. This explains the maximum in the pairing amplitude shown in Fig.~\ref{Fig:R1}(a).

One should note that we ignored the  dependence of  $\mathcal{C}_0$ on $\delta_0$ and $\Omega$ by setting it to unity in the discussion above. This approximation is based on the fact when $\delta_0\gg \Omega$ ($E_{\text{gap}} \ll\Omega$),
polaritons are predominantly excitonic and $\mathcal{C}_0 \sim 1$. This is reflected in Fig.~\ref{Fig:R1}(a) where  $\Delta_M$ essentially depends only on $E_{\text{gap}}$
for $E_{\text{gap}}\ll \Omega$. For larger values of $E_{\textrm{gap}}$, the results for different $\delta_0$ start to deviate from each other because the approximation 
$\mathcal{C}_0 \sim 1$ breaks down. In this limit, $V_{\text{ind}}$ depends on 
$\mathcal{C}_0$, which is a function of  $\delta_0$ and $\Omega$ in a way that cannot be expressed in terms of the single parameter $E_{\text{gap}}$, see also App.\ \ref{hop_app}. Likewise, when $\delta_0<0$, the dependence of the gap on the 
different parameters cannot be expressed through the  single parameter $E_{\text{gap}}$ due to fact that $\mathcal{C}_0 < 1/2$ depends on $\Omega$ and $\delta_0$.

To further demonstrate the reduced retardation effects caused by the light coupling, in Fig.~\ref{Fig:R2}(b) we plot $V_{\text{ind}}$   as a function of $\omega$ for different momenta $\bk$ both in case of excitons 
(dashed lines) and exciton-polaritons (solid lines) for $\delta_0/\epsilon_F = \Omega/\epsilon_F = 0.5$ ($E_\text{gap}/\epsilon_F=0.31$).
Here the full dependence of $V_{\text{ind}}$ on the Hopfield coefficients is taken into account. We see that introducing the coupling to light  makes $V_\text{ind}$ broader in the frequency space as well as a larger for momenta $\bk$ corresponding to $\tilde\epsilon^\text{LP}_\mathbf{k}\lesssim \omega$, in agreement with the discussion above. 
The suppression of retardation effects due to polaritons is also illustrated in Fig.~\ref{Fig:R2}(c), 
where the momentum distribution function 
$n_\bk = \langle \hat{e}^\dag_\bk \hat{e}_\bk \rangle$ of the electrons is plotted for different values of $E_{\textrm{gap}}$.
 Increasing $E_{\textrm{gap}}$ sharpens the Fermi surface
 as the frequency dependence of the diagonal self-energy  $\Sigma_{11}$ decreases as shown in App.~\ref{AppC},  thereby  enhancing superconductivity~\cite{mahan:book}.

So far we have kept the density ratio constant at $n_0/n_e = 10$. In Fig.~\ref{Fig:R3}(a), we plot the gap $\Delta_M$ as a function of the BEC density $n_0$. This shows that it depends non-monotonically on $n_0$, vanishing both for small and large $n_0$, reaching a maximum  value in between. It should be noted  that the polaritons likely are affected by the electrons when $n_0 \lesssim n_e$, which is not taken into account by 
our theory. The result of Fig.~\ref{Fig:R3}(a) showing that superconductivity is lost for $n_0 \rightarrow 0$ is however physically robust, since it is caused by the interaction mediated by the condensate being proportional to its density, i.e.\
$V_\text{ind}\propto n_0$ in Eq.~\eqref{ind_int}.
 The suppression  of superconductivity shown in Fig.~\ref{Fig:R3} for large $n_0$ can on the other hand be explained by the fact that the range of the induced interaction is determined by the BEC coherence length, i.e. $\xi \sim 1/\sqrt{2m_x(E_{\textrm{gap}} + 2n_0g_{xx})}$, see 
Appendix~\ref{AppA3}. Increasing $n_0$ thus reduces the induced interaction range leading to suppression of the Cooper pairing. Similar results regarding the density dependence of pairing were reported for atomic Bose-Fermi mixtures~\cite{Wu2016,Kinnunen2018}.

%%%%%%%%%%%%%%%%%%%%%%%%%%%%%%%%%%%%%%%%%%%%%%%%%%%%%%%%%%%%%%%%%%%%%%%%%%%
%%%%%%%%%%%%%%%%%%%%%%%%%%%%%%%%%%%%%%%%%%%%%%%%%%%%%%%%%%%%%%%%%%%%%%%%%%%

%\paragraph{Temperature dependence.--}
\subsection{Temperature dependence}
\begin{figure}
  \centering
    \includegraphics[width=1.0\columnwidth]{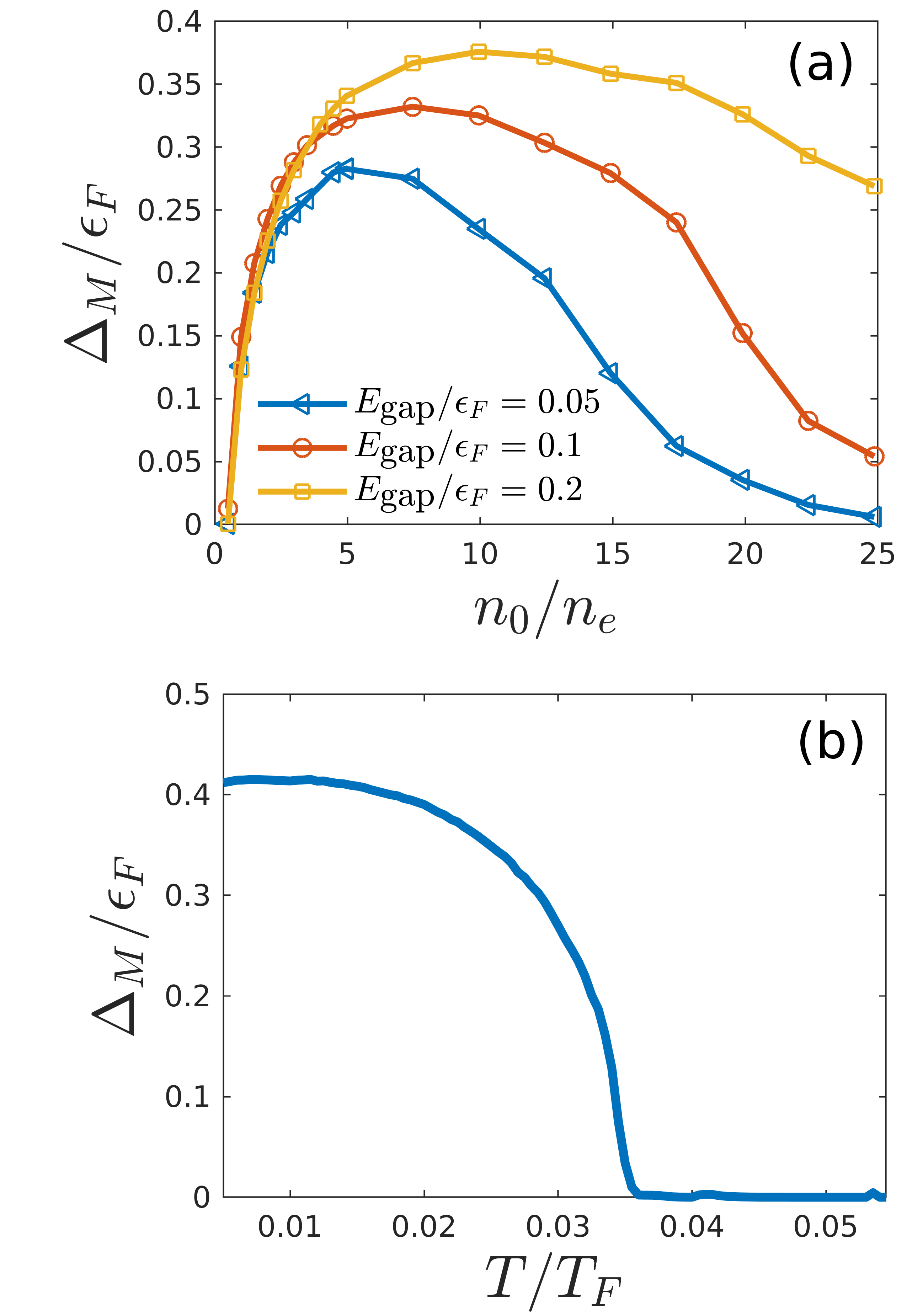}
    \caption{(a) $\Delta_M$ as a function of $n_0$ for $\delta_0/\epsilon_F = 1$, $\Omega_0/\epsilon_F = 0.23$ ($E_{\textrm{gap}}/\epsilon_F = 0.05$, blue), $\delta_0/\epsilon_F =1$, $\Omega_0/\epsilon_F = 0.33$ ($E_{\textrm{gap}}/\epsilon_F = 0.1$, red), and $\delta_0/\epsilon_F = 1$, $\Omega_0/\epsilon_F = 0.49$ ($E_{\textrm{gap}}/\epsilon_F = 0.2$, yellow). (b) $\Delta_M$ as a function of
    temperature for $\delta_0/\epsilon_F = 5$ and $\Omega/\epsilon_F = 1.37$ ($E_{\textrm{gap}}/\epsilon_F = 0.35$).}
   \label{Fig:R3}
\end{figure}

The critical temperature of the topological  superconducting phase is obviously an experimentally important quantity. To investigate this, we plot in Fig.~\ref{Fig:R3}(b) the gap $\Delta_M$ as a function of temperature for  $\delta_0/\epsilon_F =5$ and $\Omega/\epsilon_F = 1.37$ ($E_{\textrm{gap}}/\epsilon_F = 0.35$). These parameters are chosen so that $\Delta_M$ is close to its  maximum value, see  Fig.~\ref{Fig:R1}(b). This demonstrates  that the critical temperature is $T_c \simeq 0.035 T_F$, which corresponds to $T_c\sim 0.4$ K for $n_e= 10^{15}$ m$^{-2}$. Importantly, such low temperatures are within experimental reach using e.g.\ a He-3/He-4 dilution  refrigerator~\cite{Chervy2020} that  gives access  to temperatures down to  tens of mK. Note that the ratio  $T_c / \Delta_M(T=0)\sim 0.085$ is an order of magnitude smaller than the usual BCS result $0.57$. This is due to the momentum and frequency dependence of $V_{\text{ind}}$, causing the superconducting gap to be peaked at the Fermi surface as shown in App.~\ref{AppC}, and thus the Cooper pairs  more loosely bound than in case of the constant BCS gap. Other strong coupling effects such as the blurring of the Fermi surface also contribute to this effect~\cite{mahan:book}. Moreover, in contrast to the BCS result, the value of $T_c / \Delta_M(T=0)$ is not universal but depends on the values of the physical parameters.
%due to non-trivial interplay of various parameters of our system.

%%%%%%%%%%%%%%%%%%%%%%%%%%%%%%%%%%%%%%%%%%%%%%%%%%%%%%%%%%%%%%%%%%%%%%%%%%%
%%%%%%%%%%%%%%%%%%%%%%%%%%%%%%%%%%%%%%%%%%%%%%%%%%%%%%%%%%%%%%%%%%%%%%%%%%%

%\paragraph{Discussion and outlook.--} 
\section{Discussion and outlook}
We demonstrated that electrons interacting with an exciton-polariton BEC in a monolayer TMD represent a promising system to observe a 2D topological superconductor. The coupling to light was shown to  be crucial for achieving this, since it suppresses the retardation effects and makes the induced  interaction stronger in regions important for pairing. We note that this conclusion is  reliable, since it is based on a well-known expression for the induced interaction mediated by sound modes of a BEC, generalised to the hybrid light-matter case at hand. An appealing feature of our proposal is its large degree of flexibility. Experimentally, one can tune the Rabi splitting, the cavity detuning, and the densities of the BEC and electrons to optimize the range and relative strength of the pairing interaction
and thereby increase the critical temperature.

In contrast to earlier works exploring polariton-mediated superconductivity with trivial topology in multilayer setups~\cite{Laussy2010,Cotlet2016}, the  spin-valley degrees of freedom of our single layer TMD system naturally realises a 
 topological superconductor. In addition, using the 
flexibility of our system, we achieve a critical temperature 
comparable to that of Ref.~\onlinecite{Cotlet2016} despite the fact 
that our electron density is roughly one hundred times smaller.
%This low electron density also means that  our $T_c$ is not limited by a roton instability as in Ref.~\onlinecite{Cotlet2016}.

While we have used equilibrium theory to describe  the condensate, it is intrinsically in a non-equilibrium steady-state determined by the balance between a pump laser and the continuous photon loss through the cavity mirrors. This steady-state can be described using a generalised Bogoliubov theory that yields an excitation spectrum of the same form as the one used here, where the chemical potential is replaced by the frequency of the pump laser~\cite{Carusotto2013}. 
The dissipation of the polaritons due to photon leaking out of the cavity is moreover strongly suppressed for the phonon modes mediating the attractive interaction between the electrons, since polaritons are almost purely excitonic. Indeed,  their damping rate is given by $\gamma^{LP}_\bk = \mathcal{S}_\bk^2 \gamma_c$, with  $\gamma_c$ being  the photon damping rate and $\mathcal{S}_\bk^2 = 1 - \mathcal{C}^2_\bk\ll 1$ their photonic component, which is strongly suppressed
expect for very small momenta.
 Furthermore, since $E_{\text{gap}} \sim \Omega \gg \gamma_c$ for the maximal pairing gap, the effects of photon losses on the Bogoliubov spectrum are further suppressed. 
For these reasons, we  expect our results to be reliable for the pump-loss setup considered even though we have applied equilibrium theory. 

The external pump can also cause heating effects on the sample, especially if one wants to achieve high polariton densities. However, the heating effects on the electrons can be avoided by using a bilayer setup~\cite{Schwartz2021,Kuhlenkamp2021} where excitons and electrons exist in separate layers and their interaction is tuned via the Feshbach resonance. In this case the heating effects of the pump does not affect electrons, and at the same time the electron-exciton interaction can be tuned. This can furthermore be used to 
increase the critical temperature, which is an interesting topic for future 
investigations.  

Our results open up several other   new research directions  into polariton-mediated  superconductivity. This includes treating  the back-action of the electrons onto the excitons as well as the effects of superconductivity on the screening of the Coulomb interaction within the Eliashberg framework. One could also explore bilayer TMDs, where a relative twist angle produces a long-wavelength moir\'e 
lattice and flat Bloch bands~\cite{Tang2020,Regan2020,Wang2020}. One has already observed excitons in moir\'e lattices~\cite{Tran2019,Alexeev2019,Shimazaki2020,Seyler2019,Jin2019}, and flat bands give rise to strong correlations and non-trivial  superfluid 
properties~\cite{Heikkila2011,Kopnin2011,Peotta2015,Liang2017}. Furthermore, polariton-mediated ferromagnetism has been recently observed in moir\'e TMDs~\cite{Wang2022}.  In general, there are a plethora of interesting questions concerning interacting Bose-Fermi mixtures  that can  be addressed using  exciton-electron mixtures in TMDs, which will most likely complement the substantial experimental effort investigating atomic Bose-Fermi mixtures~\cite{Park2012,Heo2012,FerrierBarbut2014,Vaidya2015,DeSalvo2017,Lous2018,Schaefer2018}.

%-lifetime of intralayer excitons -> polaritons, hybrid excitons

%\begin{acknowledgments} 
\textit{Acknowledgements}--- We acknowledge very useful discussions with Ata\c{c} 
\.{I}mamo\u glu. A.\ J.\ acknowledges financial support from the Jenny and Antti Wihuri Foundation. The calculations presented above were performed using computer resources within the Aalto University School of Science “Science-IT” project. This work has been supported by the Danish National Research Foundation through the Center of Excellence “CCQ” (Grant agreement no.: DNRF156).
%\end{acknowledgments}

\appendix
\section{BEC-mediated interaction}\label{AppA}

Here we show the derivation for the induced electron-electron interaction $V_{\text{ind}}(\bk,i\omega_n)$, i.e. Eq. (2) of the main text, arising due to the exchange of sound modes of the Bose-Einstein condensation (BEC) of polaritons. The starting point is the electron-exciton interaction
\begin{align}
\label{s1}
\hat H_{\text{e-x}} = g_{ex} \sum_{\bk,\bk',\bq} \hat{x}^\dag_{\bk'-\bq} \hat{e}^\dag_{\bk+\bq} \hat{e}_\bk \hat{x}_{\bk'}   
\end{align}
where the system area is taken to be unity, $\hat{e}^\dag_{\bk}$ ($\hat{x}^\dag_{\bk}$) annihilates an electron (exciton) of momentum $\bk$ and we have assumed a contact interaction $g_{ex}$. To take into account the light-matter coupling and the emergence of polaritons, we write the exciton operator as $\hat{x}_\bk = \mathcal{S}_\bk \hat{\gamma}^{UP}_\bk + \mathcal{C}_\bk \hat{\gamma}^{LP}_\bk$, where $\mathcal{C}_\bk^2 = 1/2 + \delta_\bk/(2\sqrt{\delta_\bk^2 + 4\Omega^2})$ and $\mathcal{S}_\bk^2 = 1 - \mathcal{C}_\bk^2$ are the Hopfield coefficients and $\hat{\gamma}_\bk^{LP}$ ($\hat{\gamma}_\bk^{UP}$) is the annihilation operator for the lower (upper) branch exciton-polariton of momentum $\bk$ and energy $\epsilon^{LP}_\bk$ ($\epsilon^{UP}_\bk$). As the condensation takes place within the lower polariton branch, we can safely discard the upper polariton branch to recast Eq.~\eqref{s1} as
\begin{align}
\label{s2}
\hat H_{\text{e-x}} = g_{ex} \sum_{\bk,\bk',\bq} \mathcal{C}_{\bk'} \mathcal{C}_{\bk' -\bq} \hat{\gamma}^\dag_{\bk'-\bq} \hat{e}^\dag_{\bk+\bq} \hat{e}_\bk \hat{\gamma}_{\bk'}     
\end{align}
where for simplicity we have denoted $\hat{\gamma}_\bk \equiv \hat{\gamma}^{LP}_\bk$. In the same way, exciton-exciton interaction is now written as
\begin{align}
\label{s3}
&\hat H_{\text{x-x}} = g_{xx} \sum_{\bk,\bk',\bq} \hat{x}^\dag_{\bk'-\bq} \hat{x}^\dag_{\bk+\bq} \hat{x}_\bk \hat{x}_{\bk'} \nonumber \\
&\approx g_{xx} \sum_{\bk,\bk',\bq} \mathcal{C}_{\bk'-\bq} \mathcal{C}_{\bk +\bq}  \mathcal{C}_{\bk} \mathcal{C}_{\bk'}   \hat{\gamma}^\dag_{\bk'-\bq} \hat{\gamma}^\dag_{\bk+\bq} \hat{\gamma}_\bk \hat{\gamma}_{\bk'}.
\end{align}

As the polaritons condense at $\bk=0$, we can take $\hat{\gamma}_\bk \approx \sqrt{n_0} \delta_{\bk,0} + \delta \hat{\gamma}_\bk$, where $n_0$ is the condensate density and $\delta \hat{\gamma}_\bk$ describes the fluctuations around the BEC ground state, with $\langle \delta \hat{\gamma}_\bk \rangle =0$. In order to treat the polariton-mediated interaction, it is useful to define the 2x2 bosonic Green's function for polaritons as
\begin{align}
&G_B(\bk,\tau) \equiv -\left\langle T_\tau   \begin{bmatrix} \delta \ba(\tau) \\ \delta \bcm(\tau) \end{bmatrix}   
  \begin{bmatrix} \delta \bc(0) & \delta \bam(0) \end{bmatrix} \right \rangle \nonumber \\
  &= 
  \begin{bmatrix}
   -\langle T_\tau \delta\ba(\tau) \delta\bc(0) \rangle & -\langle T_\tau \delta\ba(\tau) \delta\bam(0) \rangle  \\   -\langle T_\tau \delta\bcm(\tau) \delta\bc(0) \rangle &  -\langle T_\tau \delta\bcm(\tau) \delta\bam(0) \rangle,
  \end{bmatrix}
\end{align}
where $T_\tau$ is the time ordering operator and $\tau$ is the imaginary time. Now, within the Bogoliubov theory, we expand the polariton-polariton interaction in Eq.~\eqref{s3} up to the second order in the fluctuation operators and ensure that terms linear in $\hat{\gamma}_{\bk=0}$ vanish by the virtue of the Hughenholtz-Pines theorem (such that the BEC is a stable ground state). Consequently, one can cast the inverse of the $2\times 2$ bosonic Green's function for the polaritons in the Matsubara frequency space as
\begin{align}
\label{g1}
&G^{-1}_B(\bk,i\omega_n) = \nonumber \\
&\begin{bmatrix}
i\omega_n - \tilde{\epsilon}_\bk^{LP} - g_{xx}\mathcal{C}^2_0\mathcal{C}^2_\bk n_0 & - g_{xx}\mathcal{C}^2_0\mathcal{C}^2_\bk n_0 \\
- g_{xx}\mathcal{C}^2_0\mathcal{C}^2_\bk n_0 & -i\omega_n - \tilde{\epsilon}_\bk^{LP} - g_{xx}\mathcal{C}^2_0\mathcal{C}^2_\bk n_0
\end{bmatrix}
\end{align}
Here $\bk \neq 0$, $\omega_n$ is a bosonic Matsubara frequency and $\tilde{\epsilon}_\bk^{LP} \equiv \epsilon^{LP}_\bk - \epsilon^{LP}_0$. The Bogoliubov excitation energies $E_\bk = \sqrt{ \tilde{\epsilon}^{LP}_{\bk}(\tilde{\epsilon}^{LP}_{\bk} + 2g_{xx} \mathcal{C}^2_0 \mathcal{C}^2_\bk n_0) }$ are obtained as the poles of $G_B$.

\begin{figure}
  \centering
    \includegraphics[width=1.0\columnwidth]{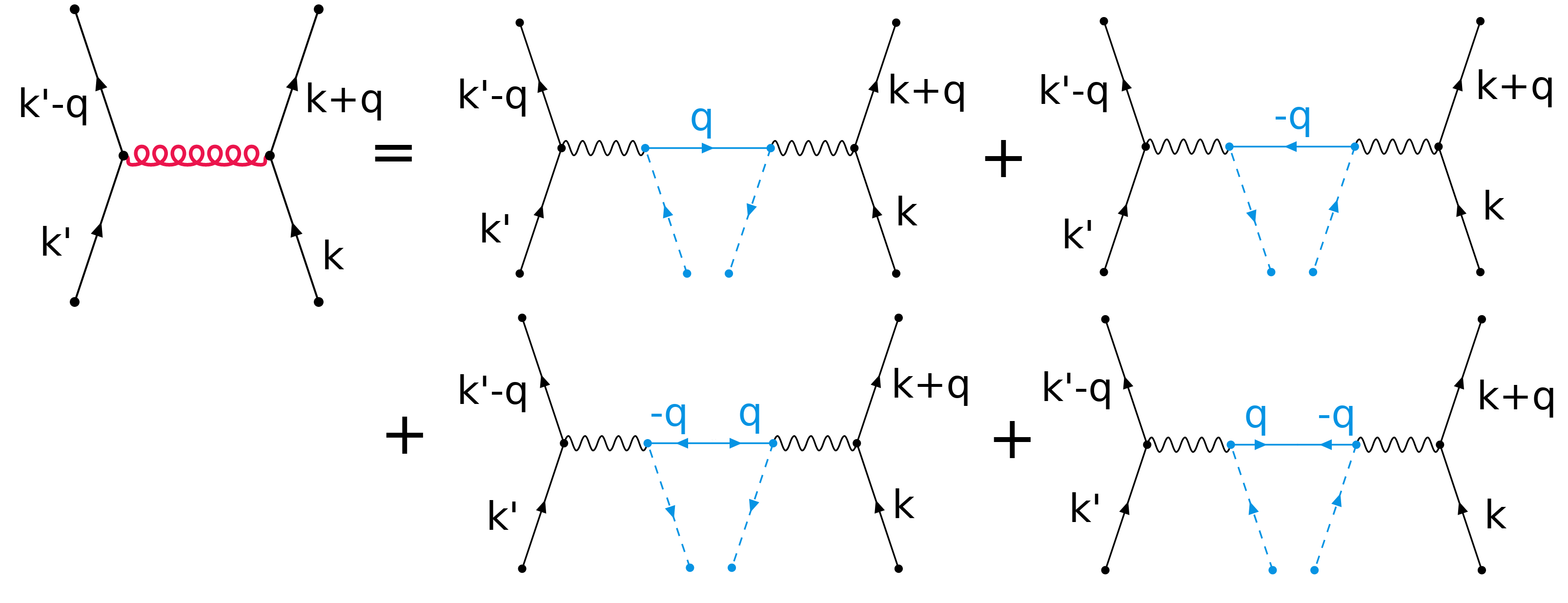}
    \caption{Feynman diagrams for induced interaction $V_{\text{ind}}(\bk,i\omega_n)$. The red curled line denotes $V_{\text{ind}}$, black wiggly lines are the bare vertex $g_{ex}\mathcal{C}_0 \mathcal{C}_\bq$, dashed blue (solid) lines are the polariton propagators of the condensate (Bogoliubov excitations) and black solid lines are electron propagators. }
   \label{Fig:S0}
\end{figure}

We can now derive the BEC-mediated electron-electron interaction. By using the form $\hat{\gamma}_\bk \approx \sqrt{n_0} \delta_{\bk,0} + \delta \hat{\gamma}_\bk$ in Eq.~\eqref{s2}, keeping the terms up to the linear order in the fluctuation operators $\delta \hat{\gamma}_\bk$ and ignoring constant shifts in the electronic chemical potential, we obtain
\begin{align}
\label{s4}
\hat H_{\text{e-x}} = g_{ex}\sqrt{n_0} \sum_{\bk,\bq} \mathcal{C}_0 \mathcal{C}_\bk \hat{e}^\dag_{\bk+\bq}  \hat{e}_{\bk} \Big( \delta\hat{\gamma}_\bq +  \delta\hat{\gamma}_{-\bq} \Big). 
\end{align}
This is of the same form as the electron-phonon coupling responsible for the phonon-mediated superconductivity. By following the usual procedure, i.e. writing down the perturbation expansion for the electronic Green's function~\cite{bruus:book}, it is straightforward to show that the interaction~\eqref{s4} leads to the effective electron-electron interaction of the form
\begin{widetext}
\begin{align}
\hat H_{\text{ind}} = \frac{1}{2}\sum_{\bk,\bk',\bq} \left( \sum_{ij}^2 [G_B(\bq,\tau)]_{ij} g_{ex}^2 n_0 (\mathcal{C}_0 \mathcal{C}_\bk)^2 \right)  \hat{e}^\dag_{\bk+\bq}  \hat{e}^\dag_{\bk'-\bq}(\tau)  \hat{e}_{\bk'}(\tau)  \hat{e}_{\bk} \equiv \frac{1}{2}\sum_{\bk,\bk',\bq} V_{\text{ind}}(\bq,\tau) \hat{e}^\dag_{\bk+\bq}  \hat{e}^\dag_{\bk'-\bq}(\tau)  \hat{e}_{\bk'}(\tau)  \hat{e}_{\bk},
\end{align}
\end{widetext}
where $\bq \neq 0$. With the Bogoliubov Green's function Eq.~\eqref{g1}, the induced interaction $V_{\text{ind}}$ can be written in the Matsubara space as
\begin{align}
\label{Vind}
&V_{\text{ind}}(\bk,i\omega_n) =  g_{ex}^2 n_0 (\mathcal{C}_0 \mathcal{C}_\bk)^2  \sum_{ij}^2 \left[G_B(\bk,i\omega_n)\right]_{ij} \nonumber \\
&= -\frac{g_{ex}^2 n_0 (\mathcal{C}_0 \mathcal{C}_\bk)^2\tilde{\epsilon}^{LP}_\bk}{\omega_n^2 + E^2_\bk}
\end{align}
which is Eq. 2 in the main text. The Feynman diagrams of $V_{\text{ind}}(\bk,i\omega_n)$ are shown in Fig.~\ref{Fig:S0} and Eq.~\eqref{Vind} can be easily obtained from them in a straightforward manner by using the standard rules for the bosonic propagators~\cite{Fetter1971}.

\subsection{Possible roton instabilities of the Bogoliubov spectrum}\label{roton_inst}

In Refs.~\cite{Cotlet2016,Matuszewski2012}, where a bilayer setup of electrons residing in one layer and exciton-polaritons in a separate TMD monolayer was considered, it was shown that the electron-hole excitations of the electron sea can lead to a supersolid instability of the Bose-condensed polariton gas, i.e. the Bogoliubov energy $E_\bk$ reaches the zero energy at finite momentum, leading to the breakdown of the assumption on the zero-momentum BEC. In Ref.~\cite{Cotlet2016} it was argued that this instability enhances the critical temperature for superconductivity in the parameter regime near the instability. As we consider electron densities much smaller than the exciton density, we expect that such an instablity does not take place in our system. Despite this and the fact that the electron densities we consider ($\sim 10^{15}$m$^{-2}$) are much smaller than those investigated in Ref~\cite{Laussy2010} ($4\times 10^{16}$m$^{-2}$) and in Ref.~\cite{Cotlet2016} ($\sim 10^{17}$m$^{-2}$), we predict a superconducting critical temperature $T_c\sim 1$K of our monolayer setup, which is of the same order of magnitude as the maximum predicted in Ref.~\cite{Cotlet2016} for the bilayer system before a supersolid instability sets in. Reasons for this difference between our and Refs.~\cite{Laussy2010,Cotlet2016} include the different momentum dependence of the electron-exciton interaction: in our monolayer setup, the excitons are tightly bound and therefore the exciton-electron interaction can be taken to be momentum-independent~\cite{Efimkin2020}. This is in stark contrast  to the bilayer system considered  in Refs.~\cite{Laussy2010,Cotlet2016}, where the  exciton-electron interaction is strongly  momentum-dependent. 

It should be noted that the random phase approximation (RPA) analysis used in  Ref.~\cite{Cotlet2016} to reveal possible supersolid instabilities would not be self-consistent in our case. Namely, the Eliashberg theory is a second order theory with respect to the exciton-electron interaction $g_{ex}$. On the other hand, the RPA renormalizes the exciton-exciton interaction as $g_{xx} \rightarrow g_{xx} + g_{ex}^2 \chi_{RPA}(\bk)$, where $\chi_{RPA}(\bk)$ is the RPA polarization bubble of the electron gas~\cite{Cotlet2016}.  If one was to include the back-action of electrons on polaritons via the RPA, then the resulting induced interaction~\eqref{Vind} would not be anymore in the 2nd order of $g_{ex}$ as the renormalized Bogoliubov energies $E_\bk$ would already include $g_{ex}$ via the renomarlization of the exciton-exciton interaction. Hence, the theory would not be self-consistent and, as a result, \textit{RPA can lead to unphysical roton minima}. To fix this, one would need to use the full Eliashberg Green’s functions in the back-action, in contrast to the ideal Fermi gas polarization bubble used  in  Ref.~\cite{Cotlet2016}. One would then proceed to compute the new Bogoliubov modes and new solutions for Eliashberg equations and continue this process iteratively till both the Bogoliubov and Eliashberg solution would converge. Such a self-consistent approach is computationally a very heavy task to implement and is thus out of scope of our work. It is important to note that the simplistic approach of Ref.~\cite{Cotlet2016} implies that the emergence of the roton minimum can actually enhance the superconducting temperature near the instability. A possibility to find roton minima in the setup considered in this work therefore  remains an interesting topic for future BEC-mediated superconductivity studies.

%Therefore, even though we discard the possibility for the roton minimum, our results for reaching superconductivity in monolayer TMDs should still be valid and actually underestimate the critical superconducting temperature.

Roton-like instabilities have been also predicted in TMD monolayer systems in Refs.~\cite{Strashko2020,Cotlet2020}. In Ref.~\cite{Strashko2020} non-zero momentum condensate states of excitons are shown to arise from the population imbalance between the holes and electrons that form excitons. In our case, there is not such a population imbalance between holes and electrons and therefore the results of Ref.~\cite{Strashko2020} do not apply in our case. The roton minima described in Ref.~\cite{Cotlet2020}, on the other hand, arise for a single exciton immersed in a Fermi sea of electrons via the Pauli blocking. This is the extreme opposite regime of what we consider, where we take the electron density to be much smaller than the exciton density.  Moreover, the electrons forming the exciton are the same as those in the Fermi sea, whereas in our study they reside in different valleys. The results of Ref.~\cite{Cotlet2020}, while interesting, are therefore not of direct relevance to our work.

\subsection{The role of the Hopfield coefficients}\label{hop_app}

The main features and behavior of the induced interaction $V_{\text{ind}}$ can be understood fairly well as a function of $E_{\text{gap}}$ by approximating $\mathcal{C}_\bk^2 \approx 1$ and  $\mathcal{C}_0^2 \approx 1$ as is done in Fig. 3(a) of the main text. For finite $\bk$ this is indeed a feasible approximation as the photons decouple from excitons at momenta that are much smaller than the ones relevant for pairing [See Fig. 2(b) in the main text]. However, $\mathcal{C}_0^2$ can in principle deviate from unity and thus affect the induced interaction strength in a non-trivial way. 

\begin{figure}
  \centering
    \includegraphics[width=1.0\columnwidth]{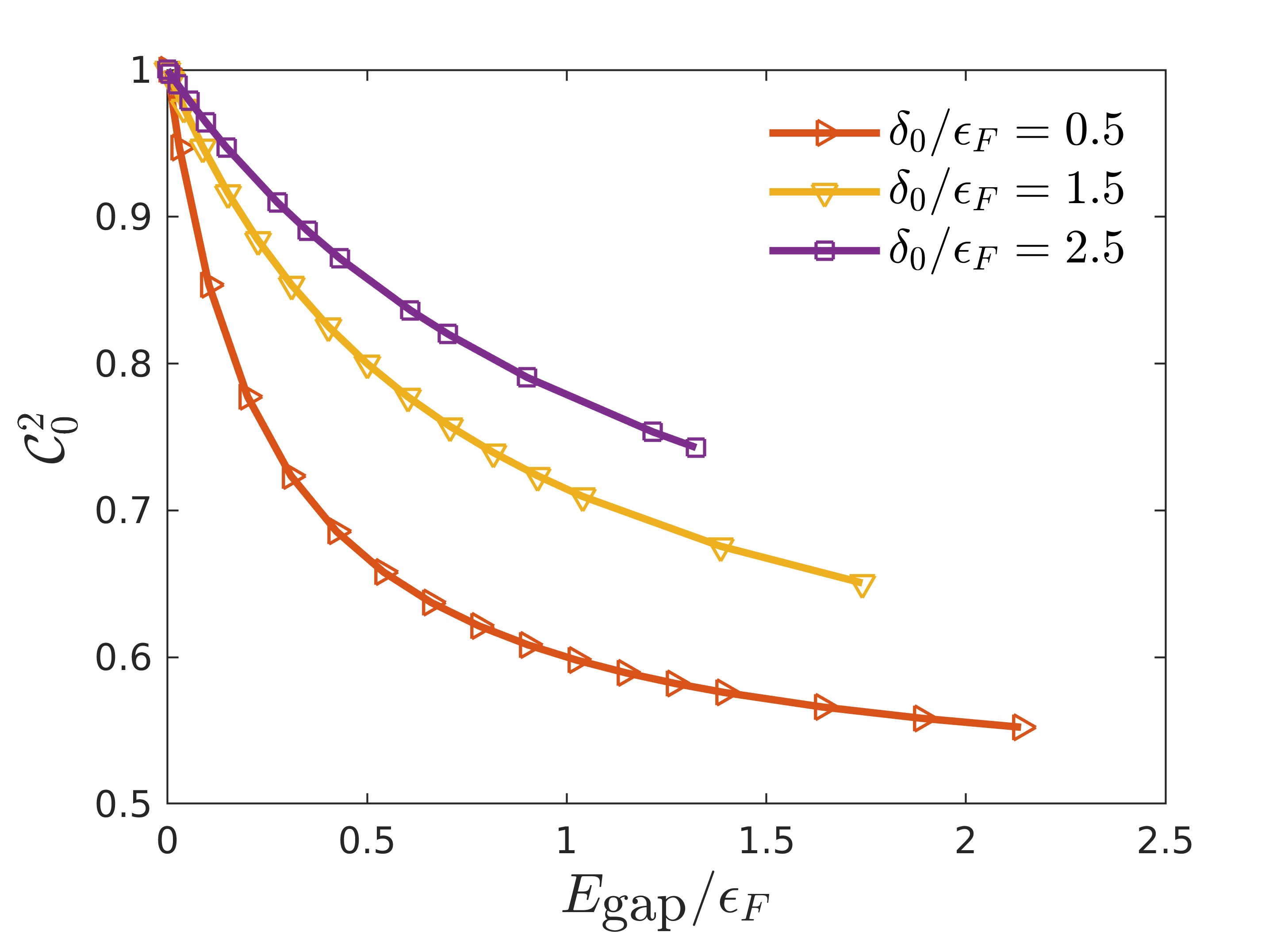}
    \caption{Hopfield coefficient $\mathcal{C}^2_0$ as a function of $E_{\text{gap}}$ for three different values of $\delta_0$.}
   \label{Fig:S1}
\end{figure}

In Fig.~\ref{Fig:S1} we plot $\mathcal{C}_0^2$ as a function of $E_{\text{gap}}$ for three different values of $\delta_0$ that were used in Fig. 2(a) of the main text. We see that for small $E_{\text{gap}}$ the Hopfield coefficients are close to unity and become gradually smaller as a function of increasing $E_{\text{gap}}$. The effect is largest for smaller $\delta_0$, consistent with the fact that smaller $\delta_0$ implies larger photonic component and thus smaller $\mathcal{C}_0$. From Fig. 2(a) of the main text we see that the pairing gap $\Delta_M$ is maximized around $E_{\text{gap}}/\epsilon_F \sim 0.25$ and all three cases of $\delta_0$ yield roughly the same $\Delta_M$. On the other hand, from Fig.~\ref{Fig:S1} we see that the corresponding $\mathcal{C}^2_0$ values are $0.75$ and $0.92$ for $\delta_0/\epsilon_F = 2.5$ and  $\delta_0/\epsilon_F = 0.5$, respectively, with a relative difference of $0.75/0.92 \sim 0.81$. Therefore, even though the Hopfield coefficients somewhat depend on $\delta_0$, the values of $\Delta_M$ in the optimal pairing regime of $E_{\text{gap}}/\epsilon_F \sim 0.25$ depend solely on $E_{\text{gap}}$. This implies that it is the reduced retardation effects, arising from finite $E_{\text{gap}}$, that dictate the pairing physics and the effect of $\mathcal{C}_0^2$ is relatively small. Only at larger $E_{\text{gap}}$ the values of $\Delta_M$ in case of different $\delta_0$ start to gradually deviate from each other (see Fig. 2(a) of the main text). From Fig. 2(a) of the main text, we see that $\Delta_M$ at the large-$E_{\text{gap}}$ regime is highest for larger  $\delta_0$, consistent with the values of $\mathcal{C}_0^2$ shown in Fig.~\ref{Fig:S1}. The behavior of $\mathcal{C}_0^2$ therefore matters only at the large $E_{\text{gap}}$ regime. As we are interested in the optimal pairing regime of $E_{\text{gap}}/\epsilon_F \sim 0.25$, the effect of  $\mathcal{C}_0^2$ can then be safely ignored when discussing the qualitative properties of the pairing interaction.

\subsection{Effective interaction range}\label{AppA3}

As we mentioned in the main text, one can approximate the polariton dispersion as $\epsilon^{LP}_\bk \approx \epsilon^x_\bk - E_{\text{gap}}\delta_{\bk,0}$. We have numerically furthermore confirmed that excluding the zero momentum does not play a role in the Eliashberg calculation so we can safely take $\tilde{\epsilon}^{LP}_\bk \approx \epsilon^x_\bk + E_{\textrm{gap}}$. With this expression, one obtains the range $\xi$ of the static interaction $V_{\text{ind}}(\bk,0)$ as ($C_\bk \approx 1$) 
\begin{align}
\label{range}
\xi \sim \frac{1}{\sqrt{ E_{\text{gap}}\big(\frac{2}{\mathcal{C}_0^2}  -1\big) +2 g_{xx} \mathcal{C}_0^2 n_0 } }.   
\end{align}
With $\mathcal{C}^2_0 \sim 1$, this is the same result as mentioned in the main text. We see from Eq.~\eqref{range} that both $E_{\text{gap}}$ and $n_0$ tend to suppress the interaction range. This can be qualitatively understood by the fact that both the quantities increase the energies of the Bogoliubov excitation modes at momenta relevant for the pairing. Thus, the exchange of the sound modes is suppressed and the range therefore decreased. 

%%%%%%%%%%%%%%%%%%%%%%%%%%%%%%%%%%%%%%%%%%%%%%%%%%%%%%%%%%%%%%%%%%%%%%%%%%%
\begin{figure*}
  \centering
    \includegraphics[width=1.0\textwidth]{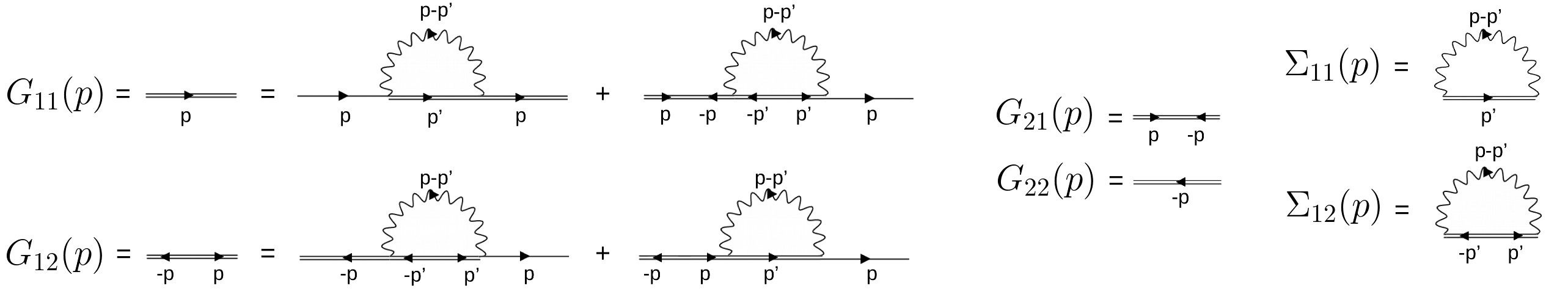}
    \caption{Feynman diagrams for $G_{11}(p)$  and $G_{12}(p)$ as well as for the self-energies $\Sigma_{11}(p)$ and $\Sigma_{12}(p)$. Single lines, double lines and wavy lines represent non-interacting propagators $G^0_{11}(p)$, interacting propagators $G_{ij}(p)$ and effective electron-electron interactions $V_{\text{tot}}(p)$, respectively. }
   \label{Fig:S2}
\end{figure*}
%%%%%%%%%%%%%%%%%%%%%%%%%%%%%%%%%%%%%%%%%%%%%%%%%%%%%%%%%%%%%%%%%%%%%%%%%%%

\section{Eliashberg equations}\label{AppB}

Eliashberg equations shown in the main text can be derived by writing down the perturbation series for the normal and anomalous Green's functions of the electrons as~\cite{bruus:book}
\begin{widetext}
\begin{align}
& G_{11}(\bk,\tau) = \frac{1}{\langle  e^{-\beta H} \rangle_0} \sum_{n=0} \frac{(-1)^n}{n!} \int_0^\beta d\tau_1 \cdots \int_0^\beta d\tau_n \langle - T_\tau H_{\text{int}}(\tau_1)\cdots H_{\text{int}}(\tau_n) \hat{e}_\bk(\tau) \hat{e}^\dag_\bk \rangle_0 \\
& G_{12}(\bk,\tau) =  \frac{1}{\langle  e^{-\beta H} \rangle_0} \sum_{n=0} \frac{(-1)^n}{n!} \int_0^\beta d\tau_1 \cdots \int_0^\beta d\tau_n \langle - T_\tau H_{\text{int}}(\tau_1)\cdots H_{\text{int}}(\tau_n) \hat{e}_\bk(\tau) \hat{e}_{-\bk} \rangle_0,
\end{align}
\end{widetext}
and keeping the terms involving the Fock diagrams~\cite{mahan:book}. Here $\langle \cdots \rangle_0$ denotes the average with respect to the non-interacting Hamiltonian, and furthermore we have defined $H_{\text{int}}(\tau) = \int_0^\beta d\tau' \frac{1}{2}\sum_{\bk,\bk',\bq} V_{\text{tot}}(\bq,\tau-\tau') \hat{e}^\dag_{\bk+\bq}(\tau')  \hat{e}^\dag_{\bk'-\bq}(\tau)  \hat{e}_{\bk'}(\tau)  \hat{e}_{\bk}(\tau')$ and $ V_{\text{tot}}(\bq,\tau) =  V_{\text{ind}}(\bq,\tau) + V_C(\bq)\delta(\tau-\tau')$ with $V_C(\bq)$ being the screened Coulomb interaction, i.e. the second term in Eq. 3 of the main text. 

As a result, one finds the diagrammatic presentation shown in Fig.~\ref{Fig:S2}. Explicitly, the Green's functions in the Matsubara space then read 
\begin{align}
&G_{11}(p) = G^0_{11}(p) + G^0_{11}(p) \Sigma_{11}(p) G_{11}(p) \nonumber \\
&+ G^0_{11}(p) \Sigma_{12}(p)G_{21}(p) \\
&G_{12}(p) = G^0_{11}(p) \Sigma_{12}(p) G_{22}(p) + G^0_{11}(p) \Sigma_{11}(p)G_{12}(p),
\end{align}
where we have used the short-hand notation $p \equiv (\bp,ip_n)$ with $p_n$ being a fermionic Matsubara frequency. The self-energies $\Sigma_{ij}(k)$ are defined as in the main text and the non-interacting Green's function reads $G^0_{11}(p) = 1/(ip_n -\epsilon^e_\bp +\mu_e)$. By using the identities $G_{21}(p) = G_{12}^*(p)$, $G_{ii}^*(p) = G_{ii}(-p)$, $G_{22}(p) = -G_{11}^*(p)$ and the fact that $V_{\text{ind}}(p) =V_{\text{ind}}(-p)$, we can write the Dyson equation as 
\begin{align}
\label{dyson}
G(p) = \begin{bmatrix}
 G_{11}(p) & G_{12}(p) \\
 G_{12}(p) & G_{22}(p) 
 \end{bmatrix} = G^0(p) + G^0(p) \Sigma(p) G(p),    
\end{align}
where $[G^0(p)]_{ij} = G^0_{ii}(p) \delta_{ij}$. By solving $G(p)$ from Eq.~\eqref{dyson}, one obtains the equations given in the main text.

\section{Momentum dependence of the gap and Fermi surface deformation}\label{AppC}

\begin{figure}
  \centering
    \includegraphics[width=1.0\columnwidth]{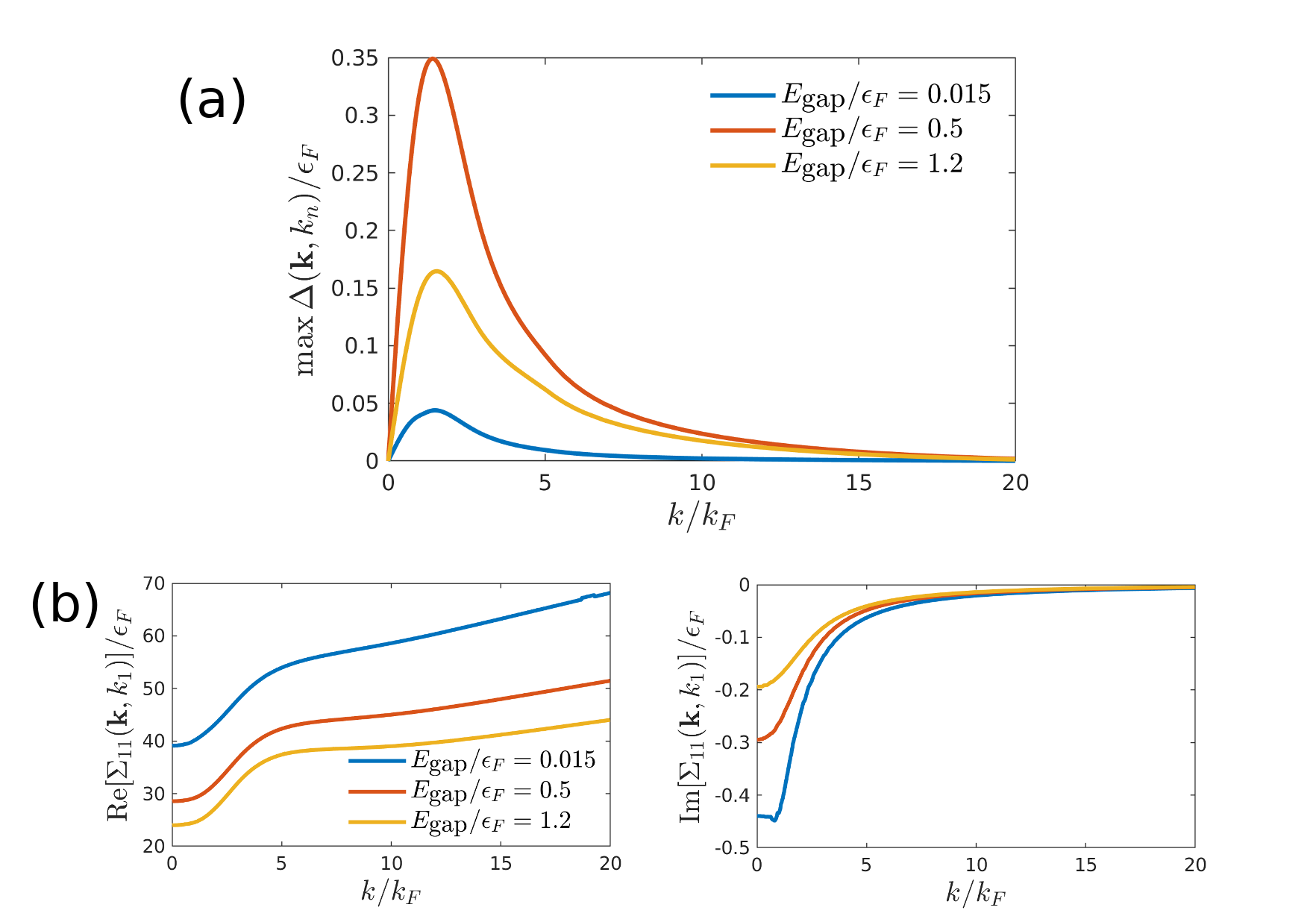}
    \caption{(a) The maximum of the pairing gap as a function of momentum $\bk$ for a few values of  $E_{\textrm{gap}}$ corresponding to $\delta_0/\epsilon_F = 1$ and $\Omega/\epsilon_F = 0.125$, $0.875$, and $1.625$. (b) Corresponding real (left panel) and imaginary (right panel) parts of the diagonal self-energy $\Sigma_{11}(\bk,k_1)$.}
   \label{Fig:S3}
\end{figure}

The superconducting critical temperature $T_c$ we obtain from our Eliashberg theory is unusually small compared to the maximum of the order parameter $\Delta_M$. The ratio of these quantities is $k_B T_c / \Delta_M ~\sim 0.044$ which is an order of magnitude lower than the BCS result 0.57~\cite{mahan:book}. However, in the usual BCS theory, the gap is assumed to be constant in the momentum and frequency spaces. Momentum-independent order parameter implies spatially small and tightly bounded Cooper pairs. In our case, the pairing gap depends strongly on the momentum as can be seen from Fig.~\ref{Fig:S3}(a), where we have plotted the maximum of the pairing gap as a function of the momentum in case of three different values of $E_{\text{gap}}$. The gap is far from being constant and is strongly peaked near the Fermi surface. Cooper pairs are therefore more loosely bound than in case of the constant BCS gap.

%%%%%%%%%%%%%%%%%%%%%%%%%%%%%%%%%%%%%%%%%%%%%%%%%%%%%%%%%%%%%%%%%%%%%%%%%%%
%%%%%%%%%%%%%%%%%%%%%%%%%%%%%%%%%%%%%%%%%%%%%%%%%%%%%%%%%%%%%%%%%%%%%%%%%%%

The BCS theory also ignores the frequency dependency of the pairing interaction. Frequency dependent interaction in turn leads to the retardation effects and finite diagonal self-energy $\Sigma_{11}(p)$. To demonstrate this in a simple way, we note that $\Sigma_{11}(p) = -\Sigma_{11}(-p)$, where we have used the properties $G_{11}(p) = -G_{11}(-p)$ and $V_{\text{tot}}(p) = V_{\text{tot}}(-p)$. Furthermore, as evidently $\Sigma_{11}(\bp,p_n) = \Sigma_{11}(-\bp,p_n)$, it is clear that $\Sigma_{11}(p) =0$ if the interaction $V_{\text{tot}}$ is frequency independent. Retarded interaction of Eq.~\eqref{Vind} therefore gives a rise for finite $\Sigma_{11}(p)$. Moreover, it is easy to see that retardation also causes the frequency dependence of the pairing gap $\Delta(p)$.

The frequency dependence of the diagonal self-energy leads to the deformation of the Fermi surface, which in turn suppresses the formation of  Cooper pairs, as the sharp Fermi surface is key to the pairing instability~\cite{mahan:book}. This manifests in Fig. 3(c) of the main text where we see how seriously deformed the Fermi surface is; the occupation number $n_\bk$ for small $\bk$ does not reach even the half-filling even though the temperature is very low ($\epsilon_F/T = 50$). By increasing $E_{\text{gap}}$, we can make the interaction $V_{\text{ind}}$ less frequency-dependent (as demonstrated in Figs. 3(a)-(b) of the main text) and in that way make the reduce the diagonal self-energy. This is shown in Fig.~\ref{Fig:S3}(b) where the real and imagainary parts of $\Sigma_{11}(\bk,k_1)$ is depicted for three $E_{\text{gap}}$ [same $\delta_0$ and $\Omega$ as in Fig. 3(c) of the main text and Fig.~\ref{Fig:S3}(a)]. We see that increasing $E_{\text{gap}}$ reduces $\Sigma_{11}$ and thus makes the Fermi surface sharper as shown in Fig. 3(c) of the main text.

\end{document}